%% file: draft_hc_to_ppbareta.tex
\RequirePackage[mathlines]{lineno} 
\documentclass[a4paper,11pt]{article}

\usepackage{jheppub} 

\usepackage[T1]{fontenc} 
\usepackage{color}
\usepackage{xcolor}
\usepackage{threeparttable}
\usepackage{multirow}
\usepackage{subfigure}
\usepackage{float}
\usepackage{ulem}
\usepackage{caption}

\renewcommand{\thetable}{\Roman{table}}  
\renewcommand{\thefigure}{\Roman{figure}}
\newcommand{\psip}{\psi(3686)}
\newcommand{\psipp}{\psi(3686)}
\newcommand{\ppp}{\pi^{+}\pi^{-}\pi^{0}}
\newcommand{\ppb}{p\bar{p}}
\newcommand{\Br}{\mathcal{B}}

\definecolor{boslv}{rgb}{0.0, 0.65, 0.58}
\definecolor{Munsell}{HTML}{00A877}
\definecolor{PurpleHeart}{HTML}{69359C}
\definecolor{ReBlue}{HTML}{333399}

\let\emph\textit
\let\oldequation\equation
\let\oldendequation\endequation
\renewenvironment{equation}
  {\linenomathNonumbers\oldequation}
  {\oldendequation\endlinenomath}

\begin{document}

\title{\boldmath Search for new hadronic decays of $h_{c}$ and observation of $h_{c}\to p\bar{p}\eta$}
\input{BESIIIauthors_BAM512}

\collaborationImg{\includegraphics[width=0.15\textwidth, angle=90]{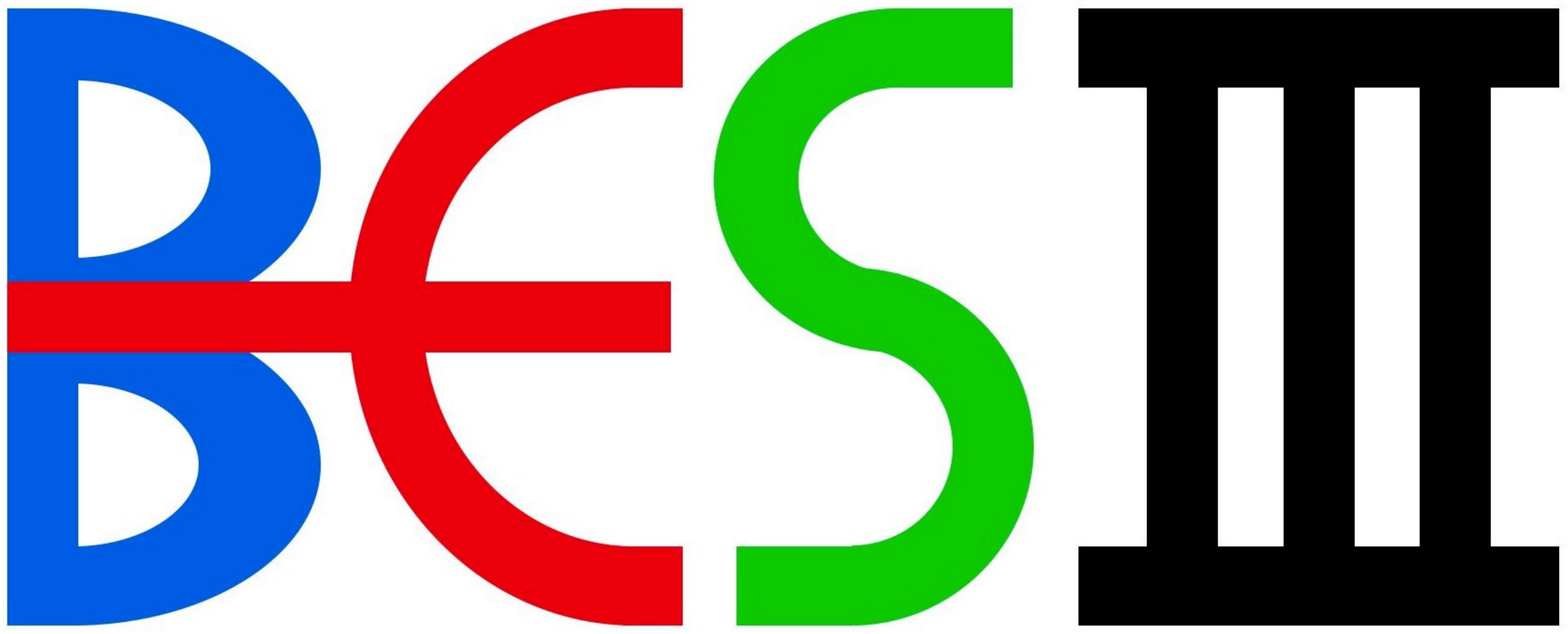}}
\collaboration{BESIII Collaboration}


\abstract{ A search for the hadronic decays of the $h_{c}$ meson to the final states
  $\ppb\ppp$, $\ppb\eta$, and $\ppb\pi^0$ via the process
  $\psipp \to \pi^{0}{h_c}$ is performed using
  $(4.48\pm0.03)\times10^{8}$ $\psi(3686)$ events collected with the
  BESIII detector. The decay channel
  $h_{c}\to\ppb\eta$ is observed for the first time with a
  significance greater than $5\sigma$ and a branching fraction of
  $\left( {6.41 \pm 1.74 \pm 0.53 \pm 1.00} \right) \times {10^{ -
      4}}$, where the uncertainties are statistical, systematic, and
  that from the branching fraction of $\psi(3686)\to\pi^{0}
  h_{c}$. Strong evidence for the decay ${h_c} \to
  p\bar{p}{\pi^+}{\pi^-}{\pi^0}$ is found with a significance of
  $4.9\sigma$ and a branching fraction of $\left( {3.84 \pm 0.83 \pm
    0.69} \pm 0.58 \right) \times {10^{ - 3}}$. The significances
  include systematic uncertainties. No clear signal of the decay
  $h_c\to p\bar{p}\pi^{0}$ is found, and an upper limit of $6.59\times
  10^{-4}$ on its branching fraction is set at the 90\% confidence
  level.

}

\maketitle
\flushbottom


\section{\boldmath Introduction}

\label{sec:intro} The study of charmonium states is crucial for a
deeper understanding of the low-energy regime of quantum
chromodynamics (QCD). All charmonium states below open-charm
$D\bar{D}$ threshold have been observed experimentally and can be well
described by potential models~\cite{Barnes}. However, knowledge about
the $P$-wave spin-singlet, $h_c(^1P_1)$, is still
sparse. Theoretically, Kuang considered the effect of $S-D$ mixing and
predicted $\mathcal{B}(h_c \to\gamma\eta_c)=(41\pm3)\%$ with a
non-relativistic QCD model~\cite{Kuang:2002hz}, while Godfrey and
Rosner predicted $\mathcal{B}(h_c \to\gamma\eta_c)=38\%$ with a QCD
model~\cite{Godfrey:2002rp}. Experimentally, the BESIII experiment
measured $\mathcal{B}(h_c \to\gamma\eta_c)=(54.3\pm 6.7\pm5.2)\%$,
which is close to these predictions, indicating that the electric
dipole ($E1$) $h_c \to\gamma\eta_c$ transition is dominant in $h_c$
decay, but that about one half of $h_c$ decays are to non-$E1$
modes. Until now, relatively few non-$E1$ decay modes, which include
two radiative decays~\cite{jielei_hc} and some light hadron
decays~\cite{cleo_hc,tracking_proton,meike_hc}, have been
observed. The sum of all measured branching fractions for $h_c$
non-$E1$ decays is $\sim 3$\%, so there are many unknown $h_c$ decay
modes after several decades of research.

The $h_{c}$ state cannot be directly produced in $e^{+}e^{-}$
collisions due to its quantum numbers ${J^{PC}} = {1^{ + -
}}$. However, it can be produced in charmonium hadronic transitions,
e.g., $\psi \left( {3686} \right) \to {\pi ^0}{h_c}$, whose branching
fraction is measured to be $\mathcal{B}(\psip \to\pi^0 h_c)=(8.4\pm
1.3\pm 1.0)\times 10^{-4}$ \cite{hc_inpsip_decay}.  In 2009 and 2012,
the BESIII experiment collected $(4.48\pm0.03)\times10^{8}$
$\psi(3686)$ events~\cite{psip:num}, which implies that about 0.4
million $h_c$ events are available via $\psip$ decays, providing a
good opportunity to study the nature of the $h_c$ state.

In this paper, we present the first search for the $h_c\to p \bar{p}
X$ ($X=\pi^{+}\pi^{-}\pi^{0}$, $\eta$, $\pi^0)$ decays via the $\psipp\to\pi^0
h_c$ process. Hereafter, we denote the three decay modes as mode
I, II, and III, respectively.

\section{BESIII detector and Monte Carlo simulation}
The BESIII detector is a magnetic spectrometer~\cite{Ablikim:2009aa}
located at the Beijing Electron Positron Collider
(BEPCII)~\cite{CXYu_bes3}. The cylindrical core of the BESIII detector
consists of a helium-based multi-layer drift chamber (MDC), a plastic
scintillator time-of-flight system (TOF), and a CsI(Tl)
electromagnetic calorimeter (EMC), which are all enclosed in a
superconducting solenoidal magnet, providing a 1.0 T magnetic
field. The solenoid is supported by an octagonal flux-return yoke with
resistive plate chamber muon identifier modules interleaved with
steel.

The acceptance of charged particles and photons is 93\% over the
$4\pi$ solid angle. The charged-particle momentum resolution at 1
GeV/$c$ is 0.5\%, and the specific energy loss ($dE/dx$) resolution is
6\% for the electrons from Bhabha scattering. The EMC measures photon
energies with a resolution of 2.5\% (5\%) at 1 GeV in the barrel (end
cap) region. The time resolution of the TOF barrel section is 68 ps,
while that of the end cap section is 110 ps. The end cap TOF system
was upgraded in 2015 with multi-gap resistive plate chamber
technology, providing a time resolution of 60
ps~\cite{tof_a,tof_b}. About 70\% of the data sample used here was
taken after this upgrade.

Simulated data samples produced with {\sc geant4}-based~\cite{geant4}
Monte Carlo (MC) software, which includes the geometric description of
the BESIII detector and the detector response, are used to determine
the detection efficiency and to estimate the background
contributions. Inclusive MC samples are produced to estimate the
contributions from possible background channels. The simulation
includes the beam energy spread and initial-state radiation (ISR) in
the $e^{+}e^{-}$ annihilations modeled with the generator {\sc
  kkmc}~\cite{kkmc_a,kkmc_b}. The ISR production of vector
charmonium(like) states and the continuum processes are also
incorporated in {\sc kkmc}~\cite{kkmc_a,kkmc_b}. The known decay modes
are modeled with {\sc evtgen}~\cite{evtgen_a,evtgen_b}, using
branching fractions from the PDG~\cite{pdg}, and the remaining unknown
decays are generated with {\sc lundcharm}~\cite{lundcharm}. Final
state radiation from charged final state particles is incorporated
with {\sc photos}~\cite{photos}.  For the exclusive MC simulation
samples, the three channels of interest are generated using the
phase-space model (PHSP) for each signal mode.


\section{Event selection and data analysis}
Each charged track reconstructed in the MDC is required to originate
from a region of 10~cm from the interaction point (IP) along the $z$
axis, which is the symmetry axis of the MDC, and 1~cm in the plane
perpendicular to $z$. The polar angle $\theta$ with respect to the $z$ axis of
the tracks must be within the fiducial volume of the MDC, $\left| {\cos
  \theta } \right| < 0.93$. The measurements of flight time in the TOF
and $dE/dx$ in the MDC for each charged track are combined to
compute particle identification (PID) confidence levels for pion, kaon
and proton hypotheses. The track is assigned to the particle type with
the highest confidence level, and that level is required to be greater
than 0.001.  Finally, a vertex fit constraining all charged tracks to
come from a common IP is made.

Photon candidates are reconstructed from isolated electromagnetic showers
produced in the crystals of the EMC. A shower is treated as a photon
candidate if the deposited energy is larger than 25 MeV in the barrel
region $\left( {\left| {\cos \theta } \right| < 0.8} \right)$ or 50
MeV in the end cap region $\left( {0.86<\left| {\cos \theta } \right|
  < 0.92} \right)$. The timing of the shower is required to be within
700 ns from the reconstructed event start time to suppress noise and
energy deposits unrelated to the event.

The $\pi^{0}$ candidates are reconstructed from $\gamma\gamma$
combinations with invariant mass within $(0.080, ~0.200)$ GeV/$c^2$. To
improve the momentum resolution, a one-constraint (1C) kinematic fit
is performed to constrain the $\gamma\gamma$ invariant mass to the
nominal $\pi^0$ mass~\cite{pdg}, and the goodness-of-fit
$\chi_{1C}^{2}(\gamma\gamma)$ is required to be less than 20. The
$\eta$ candidates are reconstructed from $\gamma\gamma$ combinations
with invariant mass within $(0.450, ~0.650)$ GeV/$c^2$, and the
invariant mass of the photon pair is constrained to the nominal $\eta$
mass~\cite{pdg} with a 1C-kinematic fit requiring
$\chi_{1C}^{2}(\gamma\gamma)<200$.

In order to reduce background events and to improve the mass
resolution, a six-constraint (6C) kinematic fit is performed
constraining the final state energy-momentum to the total initial
four-momentum of the colliding beams, and constraining the masses of
the two $\pi^{0}$s to the known $\pi^0$ mass in the $h_{c}\to
p\bar{p}\ppp$ and $h_{c}\to p\bar{p}\pi^{0}$ decays or constraining
the masses of the $\pi^{0}$ and $\eta$ mesons to their known values in
the $h_{c}\to p\bar{p}\eta$ decay. The combination with the smallest
value of the 6C-kinematic fit quality $\chi_{6C}^{2}$ is kept for
further analysis. The $\chi_{6C}^{2}$ values for $h_{c}\to
p\bar{p}\pi^{+}\pi^{-}\pi^{0}$, $h_{c}\to p\bar{p}\eta$ and $h_{c}\to
p\bar{p}\pi^{0}$ decays are required to be less than $45, 45,$ and $64$,
respectively. These values are obtained by optimizing the figure-of-merit
(FOM) defined as $S/\sqrt {S + B}$, where $S$ denotes the normalized
number of signal events, obtained from MC simulation, while $B$ is the
number of background events, obtained from inclusive MC samples.

To suppress contamination from decays with different numbers of
photons, such as the dominant decay $\psi(3686)\to\gamma\chi_{c2}$,
where the $\chi_{c2}$ decays to the same final states as the $h_{c}$,
$\chi^{2}_{4C,n\gamma}$ $<$ $\chi^{2}_{4C,(n-1)\gamma}$ and
$\chi^{2}_{4C,n\gamma}$ $<$ $\chi^{2}_{4C,(n+1)\gamma}$ are required
for each decay mode. Here $\chi_{4C,n\gamma}$ is obtained from a
four-constraint (4C) kinematic fit including the expected number of photons
$n$ in the signal candidate, while $\chi^{2}_{4C,(n-1)\gamma}$ and
$\chi^{2}_{4C,(n+1)\gamma}$ are determined from additional 4C fits with
one missing or one additional photon compared to the signal
process, respectively.

The $J/\psi$-related background is vetoed by requirements on the
$\pi^{0}\pi^{0}$, $\pi^{+}\pi^{-}$, and $\eta$ recoil masses.  The
bachelor $\pi^0$ from the decay $\psip\to\pi^0 h_c$ (identified by its
energy being closest to the expected energy) when combined with other
final state particles should not be consistent with coming from any
resonance. Therefore, additional vetoes are applied to suppress
background from $\omega\to\pi^+\pi^-\pi^0$, $\eta\to\pi^+\pi^-\pi^0$,
$\Sigma^+\to p\pi^0$ and $\bar{\Sigma}^-\to \bar{p}\pi^0$, as given in
Table~\ref{list:mass_window}, where $RM$ and $M$ denote the recoiling
mass and invariant mass, respectively, and $m$ denotes the known
mass~\cite{pdg} of the indicated particle.  The
$\Lambda/\bar{\Lambda}$-related background is also suppressed by
requiring the invariant mass of $p\pi^-/\bar{p}\pi^+$ to be out of the
$\Lambda/\bar{\Lambda}$ mass window, and the $K_S^0$ background is
rejected by requiring $m_{\pi^{+}\pi^{-}}$ to be out of the $K_S^0$
mass window. All mass windows are obtained by optimizing the FOM and
are listed in Table~\ref{list:mass_window}.  No significant
intermediate process signal is observed in the study.

\begin {table}[h]
\begin{center}
  \fontsize{8}{10}\selectfont
\begin{small}
    {\caption {Mass windows used as vetoes in each exclusive mode. }
    \label{list:mass_window}}
\begin{tabular}{c  c}
  \hline \hline

  Mode & Mass Windows [MeV/$c^{2}$] \\   \hline
  \multirow{8}*{(I)} & $\left| {RM{{\left( {{\pi ^ + }{\pi ^ - }} \right)}} - {m_{{J \mathord{\left/ {\vphantom {J \psi }} \right.  \kern-\nulldelimiterspace} \psi }}}} \right|< 15$ \\
   &   $\left| {RM{{\left( {{\pi ^ 0 }{\pi ^ 0 }} \right)}} - {m_{{J \mathord{\left/ {\vphantom {J \psi }} \right.  \kern-\nulldelimiterspace} \psi }}}} \right|<16$  \\
   &   $\left| {M\left( {{\pi ^ + }{\pi ^ - }{\pi ^0}} \right) - {m_\eta }} \right|<6$  \\
   &   $M\left( {{\pi ^ + }{\pi ^ - }{\pi ^0}} \right) \in (762,802)$   \\

   &  \small$M\left( {p{\pi ^0}} \right) \in \left( {1180,1196} \right)\&  M\left( {\bar p{\pi ^0}} \right) \in \left( {1181,1194} \right)$ \\
   &   $M\left( {p{\pi ^ - }} \right) \in (1104,1122)$  \\
   &   $M\left( {\bar{p}{\pi ^ + }} \right) \in (1104,1122)$  \\
   &   $M\left( {{\pi ^ + }{\pi ^ - }} \right) \in (490,499)$   \\  \hline
   (II) & $\left| {RM{{\left( {\eta} \right)}} - {m_{{J \mathord{\left/ {\vphantom {J \psi }} \right.  \kern-\nulldelimiterspace} \psi }}}} \right|<22$    \\   \hline
  \multirow{2}*{(III)} & $\left| {RM{{\left( {{\pi ^ 0 }{\pi ^ 0 }} \right)}} - {m_{{J \mathord{\left/ {\vphantom {J \psi }} \right.  \kern-\nulldelimiterspace} \psi }}}} \right|<30$ \\
    &      \small$M\left( {p{\pi ^0}} \right) \in \left( {1172,1202} \right)\&  M\left( {\bar p{\pi ^0}} \right) \in \left( {1171,1202} \right)$  \\
  \hline   \hline
\end{tabular}
\end{small}

\end{center}
\end{table}

After applying all selection criteria, the invariant mass
distributions for the three $h_c$ exclusive decay modes are shown in
Fig.~\ref{fit:three_Mode_mhc}. Potential background channels from the
inclusive MC sample are identified by TopoAna~\cite{zhouxy_topoAna},
which shows that the remaining background mainly originates from
resonant production with the same final state particles as the signal.
To investigate possible background from continuum processes, the
same selection criteria are applied to a data sample of
44~pb$^{-1}$ collected below the $\psi(3686)$ resonance at $\sqrt{s} = 3.65$~GeV. Only a few events
survive in modes I and III, but they are outside the $h_c$ signal region.


\section{Result}
\begin{figure}[htbp]
    \centering
    \includegraphics[width=1.0\linewidth]{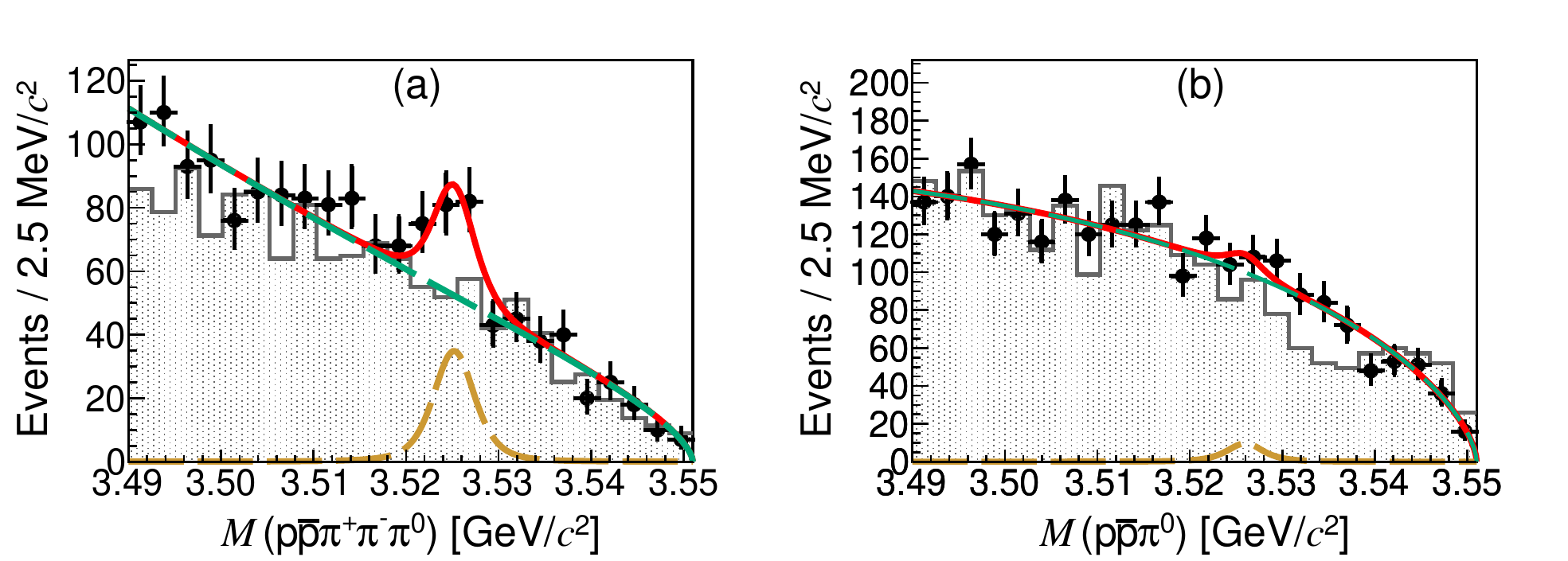}
    \includegraphics[width=1.0\linewidth]{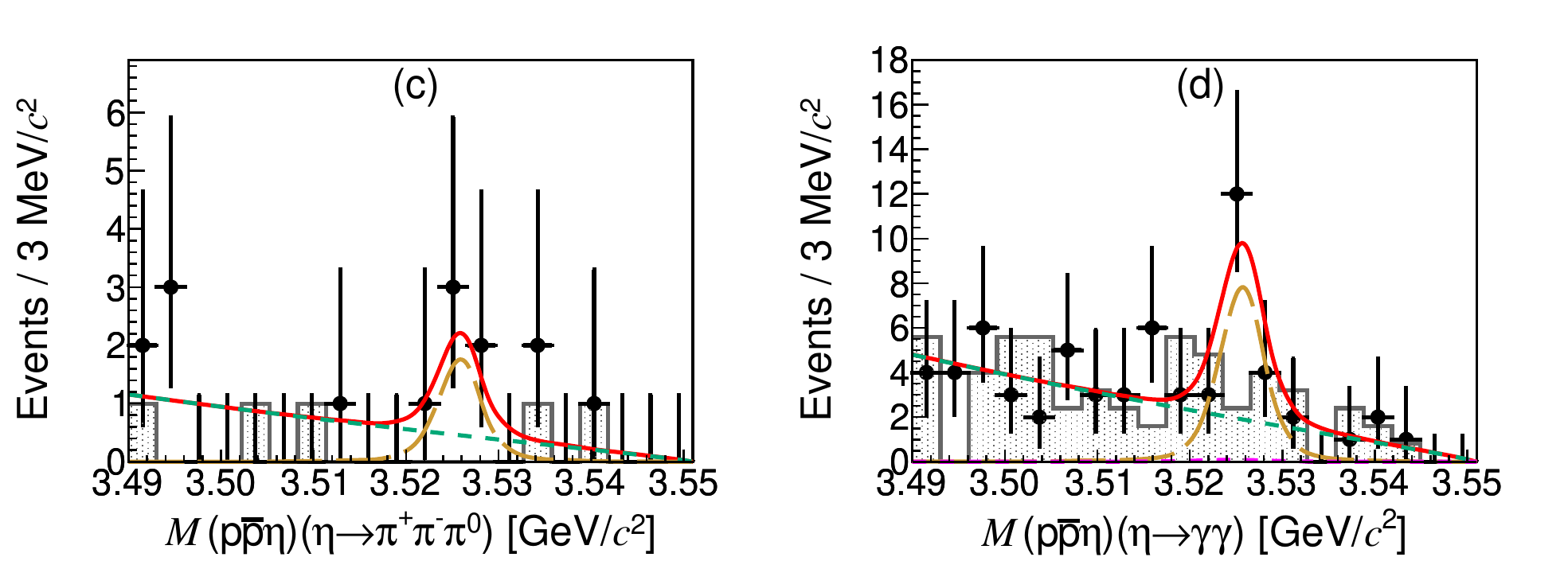}

     \caption{Fits to the invariant mass spectra of (a)
       $p\bar{p}\ppp$ and (b) $p\bar{p}\pi^{0}$ and simultaneous
       fits to the invariant mass spectra of (c) $p\bar{p}\eta$ with
       $\eta\to\pi^+\pi^-\pi^0$ and (d) $\eta\to\gamma\gamma$ for
       data. Data are shown as points with error bars, the total fit
       result is shown by the red solid line, the background contribution is
       denoted by the green dashed line (including the peaking
       background contribution shown barely
       discernible in pink in (d)), and the signal contribution is
       illustrated by the gold dashed line. The background obtained
       from inclusive MC samples is shown by the gray shaded
       histogram.  }
       \label{fit:three_Mode_mhc}

     \end{figure}

To determine the number of $h_c$ signal events $N^{\rm
  sig}$ in each decay mode, unbinned maximum likelihood fits are
performed to the corresponding mass spectra as shown in
Fig.~\ref{fit:three_Mode_mhc}.  In all fits, the signal distribution
is described by a MC-simulated shape convolved with a Gaussian
function accounting for the mass resolution difference between data
and MC simulation. The background shape is described by an ARGUS
function~\cite{Argus}, where the threshold parameter of the ARGUS
function is fixed to the kinematical threshold of
3.551~MeV/$c^{2}$. The branching fractions of $h_c\to p\bar{p}X$ are
determined by
\begin{equation}\label{determine_branch}
\mathcal{B}\left( {{h_c} \to p\bar{p}X} \right) = {\textstyle{{{N^{{\rm sig}}}} \over { {N_{\psi \left( {3686} \right) }}\cdot \mathcal{B}\left( {\psi \left( {3686} \right) \to {\pi ^0}{h_c}} \right)\cdot {\prod _{i}}  \mathcal{{B}}_{i} \cdot \varepsilon }}}.
\end{equation}
Here, {{${\prod _{i}} \mathcal{{B}}_{i}$ is the product of branching
    fractions of the decaying particles like
    $\Br(\pi^{0}\to\gamma\gamma)$ and $\Br(\eta\to\gamma\gamma)$ taken
    from the PDG~\cite{pdg}.}} The number of $\psipp$ events is determined
to be $N_{\psipp}=(448.1\pm2.9)\times10^{6}$~\cite{psip:num}. The
detection efficiencies $\varepsilon$ are obtained from signal MC
simulations, and are determined to be $6.0\%$, $18.1\%$, and $23.0\%$
for the three decay modes, respectively.

In case of mode II, there are two $\eta$ decay modes,
$\eta\to\gamma\gamma$ and $\eta\to\ppp$. A simultaneous unbinned
maximum likelihood fit is performed to determine
the branching fraction $\mathcal{B}(h_{c}\to p\bar{p}\eta)$, which is
taken as the common parameter among the different decay modes.  The
corresponding number of $h_c$ signal events in the two different
final states is calculated by:
\begin{equation}\label{eq:simufit_hc}
 {N^{\rm sig} } = {N_{\psi \left( {3686} \right)}} \cdot {\cal B}\left( {\psi \left( {3686} \right) \to {\pi ^0}{h_c}} \right) \cdot  {\cal B}\left( { h_c \to p\bar{p} \eta } \right)   \cdot {\cal B}\left( {{\pi ^0} \to \gamma \gamma } \right) \cdot {\cal B}\left( {\eta  \to X} \right) \cdot \varepsilon~.
\end{equation}
For the $\eta\to\gamma\gamma$ mode, an additional normalized peaking
background component from $h_{c}\to\gamma\eta_{c}, \eta_{c}\to
p\bar{p}\pi^{0}$ is included. For the $\eta\to\ppp$ mode, the accepted
candidate events require the invariant mass of
$\ppp$ to be in the $\eta$ signal region, i.e., $532<M_{\ppp}<562$
MeV/$c^{2}$.  The corresponding $\eta$ side-band shows no obvious
peaking background. The numerical results for $\Br(h_{c}\to
p\bar{p}X)$ and the resulting branching fraction $\Br\left(
{\psi}\left({3686}\right) \to {\pi ^0}{h_c} \right) \cdot \Br\left(
{{h_c} \to p\bar{p}X} \right)$ are listed in Table~\ref{list_summary}.

For mode III, no significant signal is observed, and an upper limit on the
branching fraction is determined by a Bayesian
approach~\cite{PhysRevD.57.3873}. To obtain the likelihood
distribution, the signal yield is scanned using the fit function,
Eq.~\eqref{determine_branch}. Systematic uncertainties are considered by
smearing the obtained likelihood curve with a Gaussian function with
the width of the systematic uncertainty of the respective decay
mode.
The upper limit at the $90\%$ confidence level on the number of events
$N^{\rm up}_{h_{c}}$ is determined by integrating the smeared
likelihood function $\mathcal{L}(N)$ up to the value $N^{\rm
  up}_{hc}$, which corresponds to $90\%$ of the integral,
\begin{equation}\label{upper_limit}
 0.9 = {\textstyle{{\int_0^{N_{{h_c}}^{\rm up}} {dN \mathcal{L}\left( N \right)} } \over {\int_0^\infty  {dN\mathcal{L}\left( N \right)} }}}~.
\end{equation}
 The results are listed in Table~\ref{list_summary} and~\ref{list_summary_B}.

Among the three $h_c$ decay modes, mode I is observed with a
statistical significance of 5.1 standard deviations ($\sigma$). The
significance for mode II is also determined to be $5.1\sigma$ by
combining the two $\eta$ decay modes, while the significance of mode
III is $1\sigma$. The statistical significance is estimated by the
likelihood difference between the fits with and without signal
component, taking the change in the degrees of freedom into account. To
evaluate the effect of the systematic uncertainty on the signal
significance, we repeat the fits with variations of the signal shape,
background shape, and fit range, and find the statistical significance
of mode II to be always larger than $5\sigma$, and mode I to be larger
than $4.9\sigma$.

\begin {table*}[htbp]
\begin{center}
\begin{small}
    {\caption {The number of observed signal events $N_{h_c}$, the absolute branching fraction ${\cal{B}}(h_c\to
        p\bar{p}X)$, the product branching fraction
        ${\cal{B}}(\psi(3686)\to\pi^0h_c)\times{\cal{B}}(h_c\to
        p\bar{p}X)$, and the statistical significance, including systematic uncertainties. Here, the first
        uncertainty is statistical, the second is systematic, and the
        third one arises from the branching fraction of
        $\psi(3686)\to\pi^0 h_c$~\cite{pdg}. }
    \label{list_summary}}
    \begin {tabular}{c c  c}\hline\hline
    Mode     & I       &    III       \\   \hline
	 $p\bar{p}X$          &   $p\bar{p}\ppp$     &  $p\bar{p}\pi^{0}$ \\
  $N_{h_{c}}$      &  $86.5\pm18.7$       &   $<57$ \\	
	   $\Br$($h_c\to p\bar{p}X$)   &      $\left( { 3.84 \pm 0.83 \pm 0.69 \pm 0.58}  \right) \times {10^{ - 3}}$  &  $<6.59 \times 10^{-4}$    \\
  $\Br$($\psipp\to\pi^{0}h_c)\times\Br(h_c\to p\bar{p}X$)  &      $\left( {3.30 \pm 0.71 \pm 0.59 } \right) \times {10^{ - 6}}$  &   $<5.67 \times 10^{-7}$     \\
Significance$(\sigma )$ &     4.9   &      $-$  \\

\hline
\hline
\end{tabular}
\end{small}

\end{center}
\end{table*}

\begin {table*}[htbp]
\begin{center}
\begin{small}
    {\caption {The number of observed signal events $N_{h_c}$, the absolute branching fraction ${\cal{B}}(h_c\to
        p\bar{p}\eta)$, the product branching fraction
        ${\cal{B}}(\psi(3686)\to\pi^0h_c)\times{\cal{B}}(h_c\to
        p\bar{p}\eta)$, and the statistical significance, including systematic uncertainties.  }
    \label{list_summary_B}}
   \begin {tabular}{c c   c}\hline\hline
    Mode       &  \multicolumn{2}{c}{ II }           \\   \hline
	 $p\bar{p}\eta$          &     $\eta\to\ppp$    &   $\eta\to\gamma\gamma$   \\
  $N_{h_{c}}$      &  $3.4\pm0.9$ & $18.1\pm4.9$    \\	
	  $\Br$($h_c\to p\bar{p}\eta$)  &  \multicolumn{2}{c}{$\left( 6.41 \pm 1.74 \pm 0.53 \pm 1.00 \right) \times {10^{ - 4}}$}      \\
  $\Br$($\psipp\to\pi^{0}h_c)\times\Br(h_c\to p\bar{p}\eta$)   & \multicolumn{2}{c}{$\left( { 5.51 \pm 1.50 \pm 0.46 } \right) \times {10^{ - 7}}$}   \\
Significance$(\sigma )$ &  \multicolumn{2}{c}{5.1}   \\

\hline
\hline
\end{tabular}
\end{small}

\end{center}
\end{table*}

\section{\boldmath Systematic uncertainty}
The sources of systematic uncertainties for the branching fractions
include tracking, photon detection, $\pi^{0}$ reconstruction, PID, the
kinematic fit, mass windows, fitting procedure, the branching fraction
of the intermediate decay, the number of $\psi(3686)$ events and the
physics model describing the $h_{c}$ production and decay
dynamics. All the systematic uncertainties are summarized in
Table~\ref{list_sys} for modes I and III, and in
Table~\ref{list_sys_2} for mode II.  The overall systematic
uncertainty for the product branching
$\mathcal{B}(\psip\to\pi^{0}h_{c})\cdot\mathcal{B}(h_{c}\to
p\bar{p}X)$ is obtained by summing all individual components in
quadrature. The third uncertainty for $\mathcal{B}(h_{c}\to p\bar{p}X)$
of ${\Delta _\text{ext}} = 15.1\% $ is due to the uncertainty of the
branching fraction of $\psipp\to\pi^{0}h_{c}$~\cite{pdg}.

\begin{itemize}
  \item {\bf{Tracking efficiency and photon detection:}} The
    uncertainties of the tracking efficiency are estimated with the
    control samples $J/\psi\to p\bar{p}\pi^{+}\pi^{-}$ and $\psip\to
    p\bar{p}\pi^{+}\pi^{-}$, and are determined to be
    $1.0\%$~\cite{uncertainty_MDC_ppi}, $1.3\%$, and
    $1.7\%$~\cite{tracking_proton} for each charged pion, proton and
    antiproton, respectively. The uncertainty of the detection
    efficiency of photons is studied using the control sample
    $J/\psi\to\ppp$, and is determined to be $1.0\%$ per
    photon~\cite{eta_recon}.
  \item {\bf \boldmath $\pi^{0}$ and $\eta$ reconstruction
    efficiencies:} There are two $\pi^0$ candidates with different
    momentum distributions for mode I and III, while there is only one
    $\pi^0$ candidate in mode II. The uncertainties of the two
    $\pi^{0}$ reconstructions differ due to the different momentum
    distributions. The corresponding uncertainties for the three decay
    modes are determined to be $0.6\%, 0.3\%, 1.8\%$,
    respectively. For the reconstruction of the $\eta$ into two
    photons, an uncertainty of $1\%$ is taken into
    account~\cite{eta_recon}.

  \item {\bf PID:} The uncertainty due to PID is determined to be
    $1.0\%$ per pion~\cite{Ablikim:092009}, $1.3\%$ per proton, and
    $1.6\%$ per antiproton, based on the same samples used to estimate
    tracking efficiencies~\cite{tracking_proton}. Tables~\ref{list_sys}
    and \ref{list_sys_2} list the relative systematic uncertainties
    due to PID for the different decay modes.

  \item {\bf Kinematic fit:} The uncertainties associated with the
    kinematic fit are studied with the track helix parameter
    correction method, as described in Ref.~\cite{uncerty_Kine}. In
    the standard analysis, these corrections are applied. The
    difference of the MC signal efficiencies without corrected track
    parameters are taken as the corresponding systematic
    uncertainties.

  \item {\bf Mass windows:} The uncertainties associated with the mass
    windows are estimated by repeating the analysis with alternative
    mass window requirements. The largest differences from the nominal
    branching fractions are assigned as the corresponding systematic
    uncertainties. In addition, the systematic uncertainties due to
    the $\Lambda/\bar{\Lambda},\Sigma^+/\bar{\Sigma}^-$ mass window
    requirements are estimated by the control samples,
    $\psipp\to\Lambda\bar{\Lambda}\pi^{0}\pi^{0}$ and
    $\psipp\to\Sigma\bar{\Sigma}\pi^{+}\pi^{-}$, respectively. The
    differences of the selection efficiencies between data and MC
    simulation from control samples are taken as the corresponding
    systematic uncertainties.

  \item {\bf Fit range:} The uncertainty due to the fitting range
    is obtained by changing the range by $\pm 0.01$~GeV$/c^{2}$,
    and the largest difference in the branching fraction is taken as
    the systematic uncertainty.

  \item {\bf Signal shape:} The uncertainty due to signal shape is
    estimated by replacing the MC-simulated shape convolved with a
    Gaussian function by only the MC-simulated shape. The difference
    in the measured branching fraction is taken as the systematic
    uncertainty.

  \item {\bf Background shape:} The uncertainties caused by the
    background shape are estimated by alternatively using different
    shapes. For mode I, we use an MC-simulated shape which is obtained
    by an MC sample of $\psip\to p h_1(1170)\bar\Delta^+ +c.c.$ and
    $\psip\to\Delta^{++}$ $\Delta^{--}\pi^{0}\pi^{0}$, to replace the
    ARGUS function.  For mode II, we use a $2^{nd}$-order Chebychev
    polynomial to replace the nominal ARGUS function.  The
    differences in the measured branching fractions are taken as the
    systematic uncertainties. The uncertainty due to
    $\eta\to\gamma\gamma$ peaking background in mode II is taken
    from the uncertainty on the branching fraction. For mode III, we
    use the ARGUS function to replace the nominal MC-simulated shape.

  \item {\bf Intermediate decays:} The maximum deviations from the known
    branching fractions are taken as systematic uncertainties as listed
    in Tables~\ref{list_sys} and~\ref{list_sys_2}.

  \item {$\bf \boldmath N_{\psipp}:$} The uncertainties due to the number
    of $\psipp$ events ($N_{\psipp}$) are determined with inclusive
    hadronic $\psipp$ decays and estimated to be
    $0.7\%$~\cite{psip:num}.

  \item {{\bf Physics model:}} The systematic uncertainties due to the
    physics model come from two sources, the unknown
    intermediate states in $h_{c}\to \text{hadrons}$, and the
    decay $\psi(3686)\to\pi^{0}h_{c}$. Since there is little knowledge
    of $h_{c}$ state decay dynamics and limited statistics in this
    analysis, we estimate the uncertainty due to possible intermediate
    resonances by including additional intermediate states, and the
    results are compared with those of the nominal phase space
    sample. For the uncertainties due to the physics model of
    $\psi(3686)\to \pi^{0}h_{c}$, we take the efficiency differences
    between the signal MC generated from PHSP and the
    HELAMP model~\cite{evtgen_a,evtgen_b} as the systematic
    uncertainties. The uncertainties for the three decay modes are determined to be $2.4\%$,
    $0.7\%$, $0.4\%$, respectively.

\end{itemize}

\begin {table}[htbp]
\begin{center}
\begin{small}
    {\caption {The relative systematic uncertainties for the $h_c\to\ppb\ppp$, $\ppb\pi^0$ decay channels (in $\%$).}
    \label{list_sys}}
    \begin {tabular}{l c c  c}\hline\hline

	Source & $h_c\to\ppb\ppp$ &  $h_c\to\ppb\pi^0$                              \\   \hline

 	Tracking     &     $5.0$      &  $3.0$                                          \\
	Photon detection       &     $4.0$     &  $4.0$                                          \\
	$\pi^{0}$ reconstruction    &    $0.6$      &  $1.8$                                         \\
    PID          &     $4.9$     &  $2.9$ 	                                 \\
	Kinematic fit   &     $1.0$       &  $0.4$                                    \\
	Mass windows       &     $12.5$     &  $7.4$                                 \\
    Fit range   	&  $5.3$     &  $1.6$                                   \\
    Signal shape	&  $4.4$      &  $3.6$                                  \\
    Background shape&  $6.8$       & $2.3$                                    \\
        Intermediate decay  &  Negligible &   Negligible                       \\
    $N_{\psipp}$  &     $0.7$      &  $0.7$            \\
   Physics model &     $2.4$      &  $0.7$            \\
	 Sum & 18.0 &  $10.6$         \\

\hline
\hline
\end{tabular}
\end{small}
\end{center}
\end{table}

Tables \ref{list_sys} and \ref{list_sys_2} summarize all the
systematic uncertainties of the different decay modes. The overall
systematic uncertainties are obtained by adding all systematic
uncertainties in quadrature assuming they are independent. For mode
II, there are two $\eta$ decay channels. Therefore, the uncommon
items are determined using the weighted average of the detection
efficiency and the branching fraction of the subsequent decays in
individual decay modes~\cite{XiaoHao_prd}.
\begin {table}[htbp]
\centering
\begin{small}
    {\caption {The relative systematic uncertainties for the $h_{c}\to p\bar{p}\eta$
        channel (in $\%$). A dash indicates that the
        systematic uncertainty is not applicable.}
    \label{list_sys_2}}
    \begin {tabular}{ l c  c}\hline\hline
	Source &    $\eta\to\gamma\gamma$  &     $\eta\to\pi^{+}\pi^{-}\pi^{0}$                   \\  \hline
 	Tracking     &     $3.0$   & $5.0$                                       \\	
	Photon detection  &                  \multicolumn{2}{c}{  $4.0$  }                           \\
	$\eta$ reconstruction     &     $1.0$   & $-$                                   \\	
$\pi^{0}$ reconstruction    &     $0.2$   & $0.6$                                         \\	

    PID          &     $2.9$   & $4.9$                                \\	
	Kinematic fit   &     $0.4$   & $1.0$                                    \\
	$\eta_{c}$ peaking background      &     $1.5$   & $-$                                   \\
   Intermediate decay &  0.5   &  1.2  \\

	veto $\psi \left( {3686} \right) \to \eta J/\psi $   &     $4.5$   & $-$                                \\
    Physics model &     $0.2$      &  $1.7$            \\
	
    Fitting range	&  \multicolumn{2}{c}{$3.7$  }                                  \\
    Signal shape	&  \multicolumn{2}{c}{$0.9$  }                                \\
    Background shape&  \multicolumn{2}{c}{$0.2$  }                                     \\
    $N_{\psipp}$  &    \multicolumn{2}{c}{$0.7$  }                                   \\
	Sum  & \multicolumn{2}{c}{ $8.3$  }                 \\    \hline
\hline

\end{tabular}
\end{small}

\end{table}


\section{\boldmath Summary}
Using a data sample of $(448.1\pm2.9)\times10^6$ $\psi(3686)$ events
collected with the BESIII detector, three decay modes of the $h_{c}$
have been searched for. The decay channel $h_{c}\to p\bar{p}\eta$ is
observed for the first time with a 5.1$\sigma$ statistical
significance, and evidence for the decay $h_{c}\to p\bar{p}\ppp$
is found with a statistical significance of $4.9\sigma$. No obvious
signal for $h_c\to\ppb\pi^0$ is seen.  The product branching
fractions, ${\cal{B}}(\psi(3686)\to\pi^0h_c)\times{\cal{B}}(h_c\to
p\bar{p}X)$, and the absolute branching fractions, ${\cal{B}}(h_c\to
p\bar{p}X)$, are listed in Table~\ref{list_summary}.  The branching fractions obtained in this
analysis are at the level of $\sim 10^{-3}$, which is the same level
as the previously observed decays of $h_c\to 2(\pi^{+}\pi^{-})\pi^0$,
$p\bar{p}\pi^{+}\pi^{-}$~\cite{tracking_proton} and $h_c\to
K^{+}K^{-}\pi^{+}\pi^{-}\pi^{0}$, $\pi^{+}\pi^{-}\pi^{0}\eta$, $K_S^0
K^{\pm}\pi^{\mp}\pi^{+}\pi^{-}$~\cite{meike_hc}.  These measurements
are essential to test the theoretical
prediction~\cite{Kuang:2002hz}. Finally, it is still unclear whether
the hadronic decay width of the $h_c$ is of the same order as the
radiative decay width predicted in~\cite{qcd_prd_1992}.  Future
experimental measurements searching for more
decay modes based on the larger data set of $\psi(3686)$
events~\cite{Future_Physic_BES3}, together with improved theoretical
calculations can help us to answer this question.

\acknowledgments
The BESIII collaboration thanks the staff of BEPCII and the IHEP computing center for their strong support. This work is supported in part by National Key Basic Research Program of China under Contracts Nos. 2020YFA0406300, 2020YFA0406400; National Natural Science Foundation of China (NSFC) under Contracts Nos. 11605042, 11625523, 11635010, 11735014, 11822506, 11835012, 11875122, 11935015, 11935016, 11935018, 11905236, 11961141012, 12022510, 12025502, 12035009, 12035013, 1204750, 12075107, 12061131003; the Chinese Academy of Sciences (CAS) Large-Scale Scientific Facility Program; Joint Large-Scale Scientific Facility Funds of the NSFC and CAS under Contracts Nos. U1732263, U1832207; CAS Key Research Program of Frontier Sciences under Contract No. QYZDJ-SSW-SLH040; 100 Talents Program of CAS; INPAC and Shanghai Key Laboratory for Particle Physics and Cosmology; ERC under Contract No. 758462; European Union Horizon 2020 research and innovation programme under Contract No. Marie Sklodowska-Curie grant agreement No 894790; German Research Foundation DFG under Contracts Nos. 443159800, Collaborative Research Center CRC 1044, FOR 2359, GRK 214; Istituto Nazionale di Fisica Nucleare, Italy; Ministry of Development of Turkey under Contract No.DPT2006K-120470; National Science and Technology fund; Olle Engkvist Foundation under Contract No. 2000605; STFC (United Kingdom); The Knut and Alice Wallenberg Foundation (Sweden) under Contract No.2016.0157; The Royal Society, UK under Contracts Nos. DH140054, DH160214; The Swedish Research Council; U. S. Department of Energy under Contracts Nos. DE-FG02-05ER41374, DE-SC-0012069; Excellent Youth Foundation of Henan Province under Contracts No. 212300410010; The youth talent support program of Henan Province under Contracts No. ZYQR201912178; Program for Innovative Research Team in University of Henan Province under Contracts No. 19IRTSTHN018.


\end{document}

%% file: BESIIIauthors_BAM512.tex
\author{\small
M.~Ablikim$^{1}$, M.~N.~Achasov$^{10,b}$, P.~Adlarson$^{68}$, S.~Ahmed$^{14}$, M.~Albrecht$^{4}$, R.~Aliberti$^{28}$, A.~Amoroso$^{67A,67C}$, M.~R.~An$^{32}$, Q.~An$^{64,50}$, X.~H.~Bai$^{58}$
% ...
\\
\vspace{0.2cm}
(BESIII Collaboration)\\
\vspace{0.2cm} {\it
$^{1}$ Institute of High Energy Physics, Beijing 100049, People's Republic of China\\
% ...
$^{4}$ Bochum Ruhr-University, D-44780 Bochum, Germany\\
% ...
$^{10}$ G.I. Budker Institute of Nuclear Physics SB RAS (BINP), Novosibirsk 630090, Russia\\
% ...
$^{14}$ Helmholtz Institute Mainz, Staudinger Weg 18, D-55099 Mainz, Germany\\
% ...
$^{28}$ Johannes Gutenberg University of Mainz, Johann-Joachim-Becher-Weg 45, D-55099 Mainz, Germany\\
% ...
$^{32}$ Liaoning Normal University, Dalian 116029, People's Republic of China\\
% ...
$^{50}$ State Key Laboratory of Particle Detection and Electronics, Beijing 100049, Hefei 230026, People's Republic of China\\
% ...
$^{58}$ University of Jinan, Jinan 250022, People's Republic of China\\
% ...
$^{64}$ University of Science and Technology of China, Hefei 230026, People's Republic of China\\
% ...
$^{67}$ University of Turin and INFN, (A)University of Turin, I-10125, Turin, Italy; (B)University of Eastern Piedmont, I-15121, Alessandria, Italy; (C)INFN, I-10125, Turin, Italy\\
$^{68}$ Uppsala University, Box 516, SE-75120 Uppsala, Sweden\\
% ...
$^{b}$ Also at the Novosibirsk State University, Novosibirsk, 630090, Russia\\
% ...
}\vspace{0.4cm}}